\documentclass[a4paper,twoside]{article}

\usepackage{epsfig}
\usepackage{subcaption}
\usepackage{calc}
\usepackage{amssymb}
\usepackage{amstext}
\usepackage{amsmath}
\usepackage{amsthm}
\usepackage{multicol}
\usepackage{pslatex}
\usepackage{apalike}
\usepackage[bottom]{footmisc}
\usepackage{graphicx}
\graphicspath{ {./figures/} }
\usepackage{xcolor}
\usepackage{SCITEPRESS}     

\begin{document}

\title{Agile methodology in online learning and how it can improve communication. A case study}

\author{\authorname{Manuela Petrescu\sup{1}\orcidAuthor{0000-0002-9537-1466}, and Adrian Sterca\sup{1}\orcidAuthor{0000-0002-5911-0269}}
\affiliation{\sup{1}Department of computer Science, Babes-Bolyai University, Str. M. Kogalniceanu, Cluj-Napoca, Romania}
\affiliation{\sup{2}Department of computer Science, Babes-Bolyai University, Str. M. Kogalniceanu, Cluj-Napoca, Romania}
\email{{manuela.petrescu, adrian.sterca}@ubbcluj.ro}
}

\keywords{Agile methodologies, online learning, educational study.}

\abstract{This paper presents a study on using Agile methodologies in the teaching process at the university/college level during the Covid-19 pandemic, online classes. We detail a list
of techniques inspired from software engineering Agile methodologies that can be used in online teaching. We also show, by analyzing students grades, that these Agile inspired techniques probably help in the educational process.}

\onecolumn \maketitle \normalsize \setcounter{footnote}{0} \vfill

\section{Introduction}


Agile is a methodology based on constant improvement that uses a different approach for software development than the classical waterfall approach. The waterfall methodology has a set of advantages such as using a clear development structure, having a goal determined in the first phases. However waterfall methodology has some major disadvantages (e.g. changes are difficult to implement, the client is excluded from the development process, so on) and because of these, it is slowly replaced by other methodologies (such as Agile). Agile sticks to some base principles in order to improve communication, cooperation, customer feedback and to ease the implementation of changes in the developed application. The main principles are \cite{1}: 
\begin{itemize}
    \item Individuals and interactions over processes and tools 
    \item Working software over comprehensive documentation
    \item Customer collaboration over contract negotiation, Responding to change over following a plan 
\end{itemize}
For a more complete presentation of Agile methodologies, please see \cite{16}.


As higher education enrollment is seeing a decline because of high tuition costs, online education could help students start and finalize their studies, a swift change in learning methods seems to be happening, going to a more ecological and economically viable environment \cite{6}. The overall post secondary enrollment rates dropped and the rates for online enrollment grew year after year \cite{4}. These directions were accentuated due to the 2019 pandemic, as in fall 2020 the undergraduate enrollment had a 4\% decline compared to the previous year, and the postsecondary enrollment had a 3\% decline compared to the previous year \cite{5}.  A comparison between 2021 and 2020 confirms the trend, spring 2021’s undergraduate enrollment decline was 4.5\% compared to 2020 \cite{2}. A market analysis performed by Market Research, found out that in 2019, the US Academic E-Learning market size was around US\$ 1.84 billion, and their prognostic was to reach US\$ 5.31 billion by the end of 2026.\cite{9}


COVID-19 pandemic obliged schools to go online, most colleges and universities moved their courses online and closed the campuses. Some students managed to attend online classes, others had problems related to internet access, mobile devices and even with finding quiet learning locations. In the USA,  a survey of college students done after the Spring 2020 semester indicated that 43\% of the students that have enrolled in traditional face-to-face classroom courses did not take an online class before. 21\% if the students had only taken one online class prior to the pandemic, and 35\% of them had taken two or more classes. \cite{2}

Even if the students and the professors seem confident in achieving the learning goals, the results point out a different thing: the students and (their parents) have a lower degree of satisfaction regarding the online tutoring comparing to traditional face-to-face learning. In USA, a study done after one online semester shows that the satisfaction level decreased from 87\% to 59\% when the courses were held online and the high dissatisfaction level grew from 3\% to 13\% \cite{2}. In another study from Romania, a significant percent of parents (23.03\%) considered that the online courses did not help their children (middle school, high school, and university level) to accumulate new knowledge.\cite{3} 
\\
\\
\underline{Pro and cons of online learning vs traditional learning}\\
\\
There are lots of factors that can be considered as pro arguments for online learning: the cost, the availability of the course content (24/7), and the availability as each student can learn at his own pace. The cons arguments are referring to less human interaction, the increase of zoom fatigue due to lack of nonverbal communication, weariness, more distractions and environmental issues (hardware/software capacity, bandwidth, Internet issues, difficulty of finding a quiet place to learn, so on). 
\\
\\
\underline{Online learning advantages}\\
\\
There are countries where the universities offer traditionally free degree programs and the students do not pay for their tuition. However, some traditional universities have costs with managing property, buildings, staff, costs that can get up to  \$30,000 to \$50,000 just in tuition for universities such as Harvard, Duke or Yale \cite{7}. These costs are largely not present for the online courses, the average cost spent  for online courses in the USA was  between \$100 to \$400 \cite{7} per credit hour and 79\% of the students that completed an online course strongly agree that it is worth the cost \cite{8}. For corporations, up to 60\% of total training cost is due to traveling expenses according to a KPMG paper \cite{10}. Other arguments are referring to the course availability, the online courses can be accessed when the students have time.
\\
\\
\underline{Online learning disadvantages} \\
\\
Taking online courses by necessity as the schools closed, has a huge impact on the students, as some of them struggle with hardware constraints (not having a mobile device to connect), to poor infrastructure, or finding a quiet space to learn. “The main reason  was the lack of technical resources to access the platform or Internet connection at home. From the students’ perspective, 81\% of those who declared that they have not participated in online courses at all, are from rural areas and do not have an Internet connection at home.” \cite{3}. In the USA, most of the students complained about the home schooling conditions as 20\% of college students who had online classes instead of traditional classes during the pandemic indicated that it was a major challenge to find a quiet place for learning \cite{2}. 
Spending more than 5, 6 hours every day in meetings has another psychological impact called “zoom fatigue”, making the students tired as their brain tries to compensate for the lack of nonverbal behaviour, effect analysed in different papers \cite{11,12}.

\section{Using Agile in online and offline learning}

There are many papers in the industry discussing how to teach the Agile methodology \cite{13} \cite{14,15}, but the literature is not so abundant in papers related to how Agile can be used when teaching a course, especially regarding the online courses. 

In this study we intend to answer to the following research questions: 
\newline
1. \emph{Is the Agile methodology helpful in the online courses?}
\newline
2. \emph{Should Agile methodologies be applied to courses /seminars or laboratories?}
\newline

Agile is a software development methodology that is taught  at college level, but the Agile concepts can be adapted to online teaching. Our proposal is to adapt the following events:
\begin{itemize}
    \item Sprint planning
    \item Sprint review
    \item Sprint prospective
    \item Agile stand-ups
    \item Use Agile for nonverbal communication.
\end{itemize}

\subsection{Scrum - Sprint planning event when teaching}

Mentioning the requirements, the syllabus on the first course, mentioning the plan and the structure for each seminar/course or laboratory in the beginning is similar to Scrum, Sprint Planning event. In each sprint two major questions would be answered: Why is the sprint valuable? and What can be done in this sprint? Following this approach, each professor can mention and answer similar questions  in the beginning of a course in order to get the interest of his students:

\begin{itemize}
    \item What it will be discussed in the course/seminar; the content and the structure. 
    \item Why the discussed topics are important
\end{itemize}

Mentioning the plan, the structure and the relation between the concepts makes the course easier to follow and to understand. Every course should be well structured, and it’s important that this information and the topics that will be discussed are shared with  the students, as it helps students understand and integrate the presented information into patterns, making the concepts easier to retrieve and to use. Every course should be well structured, starting with the plan, then presenting the information in a logical manner, and finally exercises and conclusion.  Using this approach, the students will know when it ends, how information is related and structured and what is expected from them.

In a study involving students from Babes Bolyai University, Computer Science that was performed by us (details of the study will be provided later on), most of them 97.1\% appreciated that it was useful to find out in the beginning of each seminar/laboratory/course the structure and the main topics.

\subsection{Scrum - Sprint Review when teaching}

The sprint review is a working session that analyses what was accomplished in the sprint, a session that should be time-boxed to a maximum of 4 hours for a four weeks sprint. Related to a course of two or four hours, it should be time-boxed to 5 to 10 minutes in which a summary of the major topics, the essential terminology and concepts are presented. Summarizing at the end of the course helps students remember the information presented in the higher percentage and it also helps structuring it into patterns. In our study, most of them appreciated that summarizing the information at the end of a course was helpful. Summarizing should also include a Q\&A short session. Asking questions, involving the students to participate and analyse or debate over the presented topics not only helps them clear some misunderstandings, but also has a beneficial impact on the students/professor communication. The teacher can give positive reinforcements to the desired behaviour and to interesting or challenging questions. The students, as all human beings, usually are motivated by the teacher’s appreciation, thus improving communication. The following image reflects the percentage of students in our study that appreciated having a review after the course (97.1\% appreciated a review is useful).

\begin{figure}[h]
\includegraphics{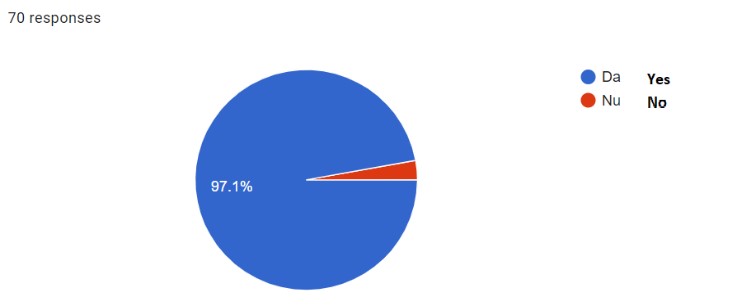}
\caption{Student's feedback regarding Sprint Review }
\label{fig:method}
\end{figure}

We made the following observation: the percentage of the students that appreciated Sprint Review was the same with the percentage of the students appreciating Sprint Planning. We also analysed that there is a possibility that the students just clicked without reasoning as the questions were displayed one after another, but we considered that the percentage is extremely high and is relevant for our study, even if some students might have been influenced by the order of the questions.

\subsection{Scrum - Sprint Retrospective Event when teaching}

The sprint retrospective is a meeting that formally concludes the sprint, in which the team discusses what went well in the sprint, what could be improved and what will be committed to improve. Normally, a sprint retrospective in Agile software development lasts anywhere between 30 minutes to 3 hours. In teaching, the duration of this retrospective/feedback meeting/discussion should be scaled down proportionally to the duration of the "teaching sprint" and we suggest is should not take more than 10 minutes and should be done after each major topic (that includes 2 - 4 course hours). This feedback discussion should not be a Q\&A session, but rather a session for gathering information about how well the students understood the explained concepts and how well they can apply them, their impressions and feelings as suggestions for methods to improve the following courses can also be collected. There are cases when the examples provided by the professors are not enough or they are not explained in detail, or there is not enough/too much theory. Knowing the student’s opinion, the professors could improve their teaching methods and the structure of the course. 
However, the students' opinions should not prevail over the overall scope of the course.

Our recommendation is not to use this instrument too frequently as asking every time for feedback the students will get bored, disregard the requirement and provide irrelevant answers or even consider that the professor is not confident enough on his capabilities and need constant feedback. However, asking the feedback only once in a semester, at the end of it, might prove not enough as the students tend to forget the observations they had in the first courses. 
 
\subsection{Agile Stand-ups meeting in teaching}

In Agile methodology, a stand up stands for a daily 15 minutes meeting for the team to plan for the next 24 working hours, and is based on three questions: What did you do yesterday? What will you do today? Anything blocking your progress? 
\\
Agile recommends that a team has a maximum 8 members, so a 15 minute meeting is feasible if the additional topics are discussed. However, due to the fact that the number of students attending a course is much higher, we recommend the following approach only for seminars or for laboratories where the student’s number is smaller. Even though the number of students in a seminar group or laboratory group is usually not as low as 8 (typically it is between 20 and 30 at our university), we consider that 15 minutes stand-ups is a good compromise between using the seminar/laboratory time for presenting new knowledge to students and discussing issues related to already acquired knowledge.
Our proposal is to have a 10-15 minutes discussion about the problems the students encountered when doing their homework and about the solutions they found. Most of the time, during the laboratories, the professor can help/check homework only for a student at a specific moment of time. Having this session in the beginning of a class has some major advantages:
\begin{itemize}
    \item All the students find new information and learn from their colleagues; discussing mistakes and problems as the solution for those problems is increasing their experience.
    \item The students become confident, opened with their colleagues and learn to work in a team. They realise that everyone is making mistakes, everyone encounters issues that have to be solved and the most important thing is to learn. 
    \item The professor can find out where there are the major problems and how to improve the course by providing other examples or details.
\end{itemize}

\section{Agile in nonverbal communication in online courses}

This semester, due to the pandemic all courses were held online. Based on the following arguments, we tried to make students have their camera turned on. The arguments were:
\begin{itemize}
    \item They become aware that they can be asked to answer and that they are seen if they do something else or are not paying attention.
    \item The students have to attend the seminars/laboratories, they are not allowed to turn off their cameras while they are leaving the room or they are getting involved in other activities.
    \item The professor can see if the students understand the explained concepts based on their face mimics and the professor can adjust the course accordingly.
\end{itemize}
As in Agile methodology, when the students have the camera turned on, the “fatigue” signs such as yawing or texting on the phone, or bending over the desks are easy to be noticed (for a small number of participants) and the professor can react and try to animate the course. There are several methods that can be used to animate a course, from joking to asking questions or starting some free discussions. 

In the 8th week of study, after applying some Agile methods in teaching, we decided to see what was the perception of the students regarding the new approach. So we create an anonymous quiz with 10 closed questions (8 with yes/no option and 2 with more than 4 multiple response options) and one open question to allow them to express other opinions/ideas related to the course and/or Agile methods that were used during this semester. In the next section, we will focus on the perception for cameras turned on/off and on the methods to animate the course.
\\
\\
\underline{Camera ON/OFF} \\
\\
In the 8th week of study, the students were required to provide a feedback for this approach. As expected, most of them mentioned that they would prefer to have their camera turned off, and in the same ratio, they wanted the professor to have the camera turned on.
Our expectation was that the ratio of 61.4\% of students that prefer the teacher to have their camera turned on was even higher, but the results should take into consideration the fact that some parts of the laboratories consisted in presenting home-works or explaining concepts/ PowerPoint presentations that imply screen sharing and decrease the importance of non verbal communication in those specific time-frames.

\begin{figure}[h]
\includegraphics [width=0.5\textwidth]{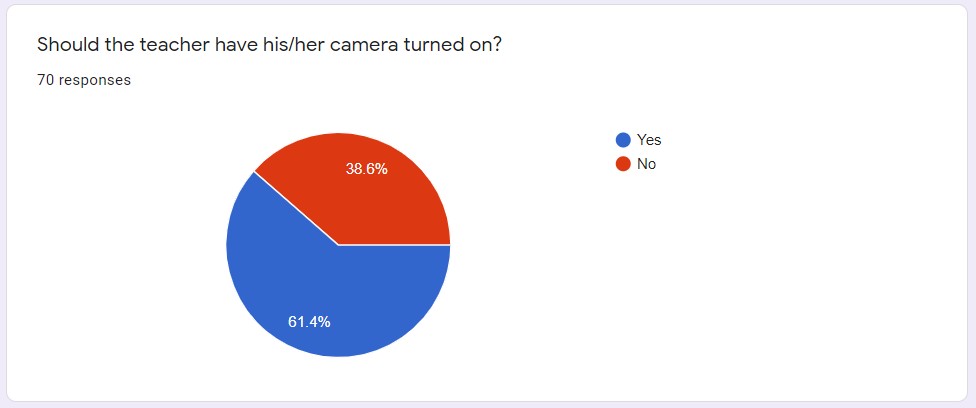}
\centering
\caption{Student's feedback regarding teacher's camera: ON/OFF }
\label{fig:method}
\end{figure}

\begin{figure}[h]
\includegraphics  [width=0.5\textwidth]{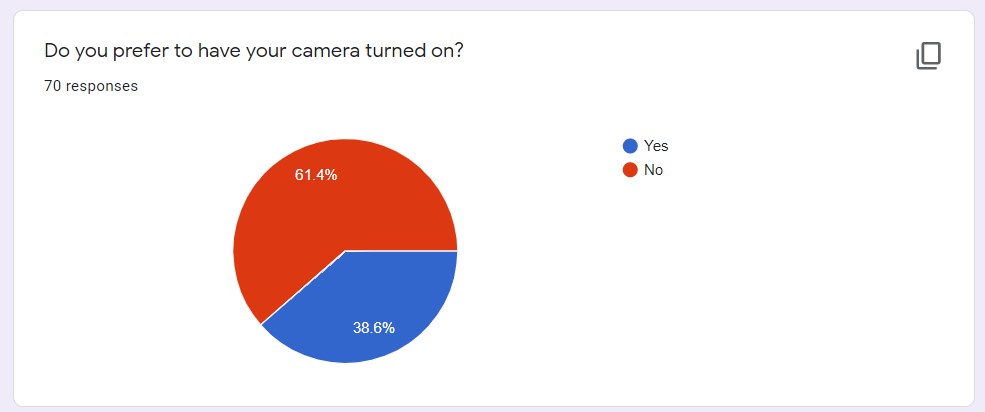}
\centering
\caption{Student's feedback regarding their own camera: ON/OFF }
\label{fig:method}
\end{figure}

As a note, even if the responses collected from students mentioned that they prefer to have their camera turned on, when they were allowed to choose to have their camera turned on/off, most of them turned off their cameras. 

As the results were surprising (38,6\% of students stated that they prefer to have their camera turned on), we decided to perform some tests as in our  observations in other groups of students the percentage was much lower. We selected 3 semi-groups of 15 students and they were allowed to choose freely to set their camera on or off. Out of this set of students only 2-3 students turned their camera on. Furthermore, if the
professor would leave his/her camera off, the students are not encouraged to turn on
their camera; the best observed ratio of students turning on their camera voluntarily was less than the declared one by 20\% - 25\%.

Another question was related to the relation between the students' involvement  in the course and their turned on/off camera. Our assumption was that the students with the camera turned on (due to the fact that they know that are observed), will be more involved. However, the percentage of students that declared that they got more involved is lower than expected: only 15.7\%, and most of them appreciated they were involved in the same manner. 

\begin{figure}[h]
\includegraphics  [width=0.5\textwidth]{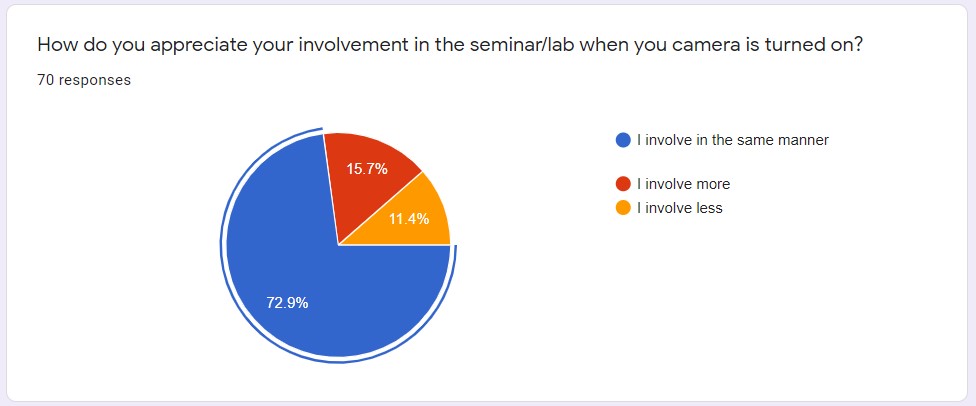}
\caption{Student's feedback regarding their involvement}
\label{fig:method}
\end{figure}

However, having students with their camera turned on was a valuable resource as the signs that showed boredom were easier to spot, so the teacher was aware and could adapt. Some students that were texting on cell phones were asked to answer, or to provide an opinion about the discussed topic. 
\\
The results of the quiz were surprising as we expected to have a higher percentage of students that get more involved if they have their camera turned on, but we did not perform additional tests to really measure their involvement (the metrics could be the number of times they respond freely without being questioned by the professor). The observed pattern was that the students are more aware of the fact that they are in a class, they try to hide themselves when texting; so in our perception, their behaviour changed when they had the cameras turned on. \\

\underline{Methods to animate the course}
\newline

Due to many distractions available at home, the students are easier to get distracted and to lose their focus. The professors need to reinvent themselves and reinvent the teaching methods in order to keep their students interested. In our study we tried different approaches: humor (jokes), asking the students that are distracted to answer, or provoking those students that have their camera turned off and free discussions.

\textbf{Using humor} has some advantages: it’s relatively easy to integrate, it consumes little time, relaxes everyone and makes the students feel better. How the students are feeling in a class is quite important as the information presented can be retrieved from books, from papers, sometimes having a better structure than the one presented at course. However, what makes a course unique, so the students love it and learn from it is the communication and the connection they are having with the professor. Humor can create a communication bridge. The downside is that when there is no connection between the students and the professor and only the professor finds the jokes amusing, the humor should not be used. It just makes communication harder.
\\

Even if humor is one of the best methods to animate a course, it has to be used with caution as no one wants to transform a course into a stand-up comedy. The main focus is and should be learning and passing information and knowledge to students.

\textbf{Provoke distracted students} to answers was a method used in our experiments. Even though the students were more attentive, they did not like this approach. They felt uncomfortable as they had to control their distractions. The seminar request was that everyone should have the camera turned on. Provoking students that had their camera turned off to answer usually had a positive effect on the other students, as the inattentive students stuttering amused their colleagues.
None of these methods can however be used intensively, as asking and provoking students that don’t know the subject or that do not pay attention distract the other students and make them lose their focus. The logical explanations and flow can also suffer, so the recommendation is to use each method in an appropriate manner.

\textbf{Free discussions} was in our opinion the most difficult method to use as it consumes time and it’s not easy to find an interesting provoking topic from the course topics. We realize that this may be different for other learning setups and may be too specific to our own experience.
A downside for using free discussions is that they can be used only after the theoretical part is taught and explained, only after the basic knowledge on a topic is achieved. However, the free discussion method was the most appreciated method for animating the course in our study, as the students can express their ideas and their personality, can argue and debate over a subject or a topic.

\textbf{Examples} were also appreciated by the students. It was interesting to find out that our assumption that the waw examples (“the most”, “the top”) would be more appreciated proved to be wrong, the students preferred familiar examples in the same percentage as waw examples. The percentage was 38.6 familiar examples to 34.3 waw examples.
\begin{figure}[h]
\includegraphics  [width=0.5\textwidth]{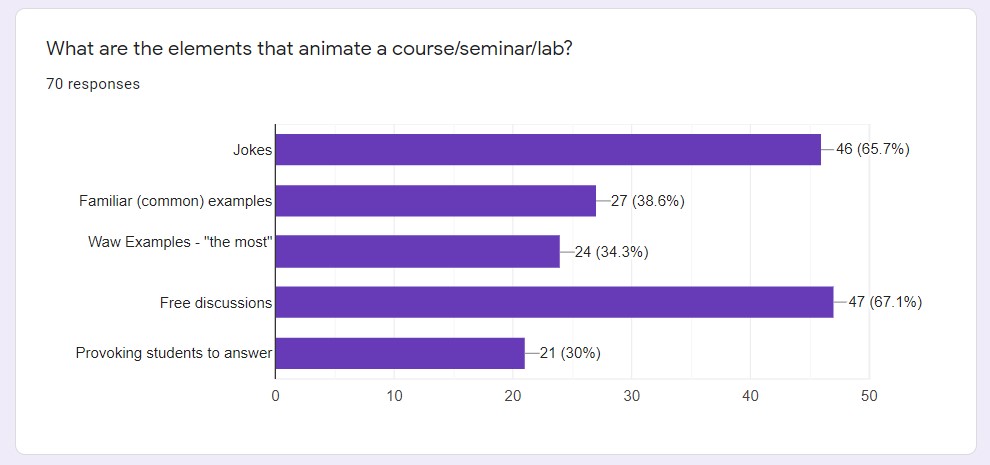}
\caption{Student's feedback regarding animation elements}
\label{fig:method}
\end{figure}

\section{Evaluation}

In this section we are trying to evaluate the effect of applying some of the above Agile techniques to Web Programming laboratory classes held with university students. At the Web Programming laboratory classes students are assigned tasks from the various fields taught at the Web Programming course (Html, CSS, Javascript, Jquery library, Angular framework, PHP programming, JSP and Java servlets, ASP.Net Core) and they have specific deadlines for each task. The grade received by a student for a lab task reflects the quality of the solution and the fact that the student complied with that task's deadline; the grades vary from 1 (received for a lab task not delivered) to 10 (received for a good to perfect solution for the lab task). 

We are comparing the lab grades received by students in academic year 2019, before the Covid-19 pandemic, when the classes were held with students being physically present in laboratory rooms, with the grades received by students in academic year 2020, during the Covid-19 pandemic, when all classes were taught online using Microsoft Teams videoconferencing sessions. In the academic year 2020 we applied at the Web Programming laboratory classes some of the Agile techniques mentioned above, as a way to help students cope with the non-physical presence at the university induced by the Covid-19 pandemic.
Since we normally expect lab grades to drop during Covid-19 pandemic with online laboratory
classes, we are trying to measure whether Agile techniques can be used in order to maintain
or improve lab grades for pandemic online classes with respect to normal, pre-pandemic, physical classes (i.e. when students were required to be physically present at the laboratory class)

\begin{figure}[h]
\includegraphics [width=0.5\textwidth]{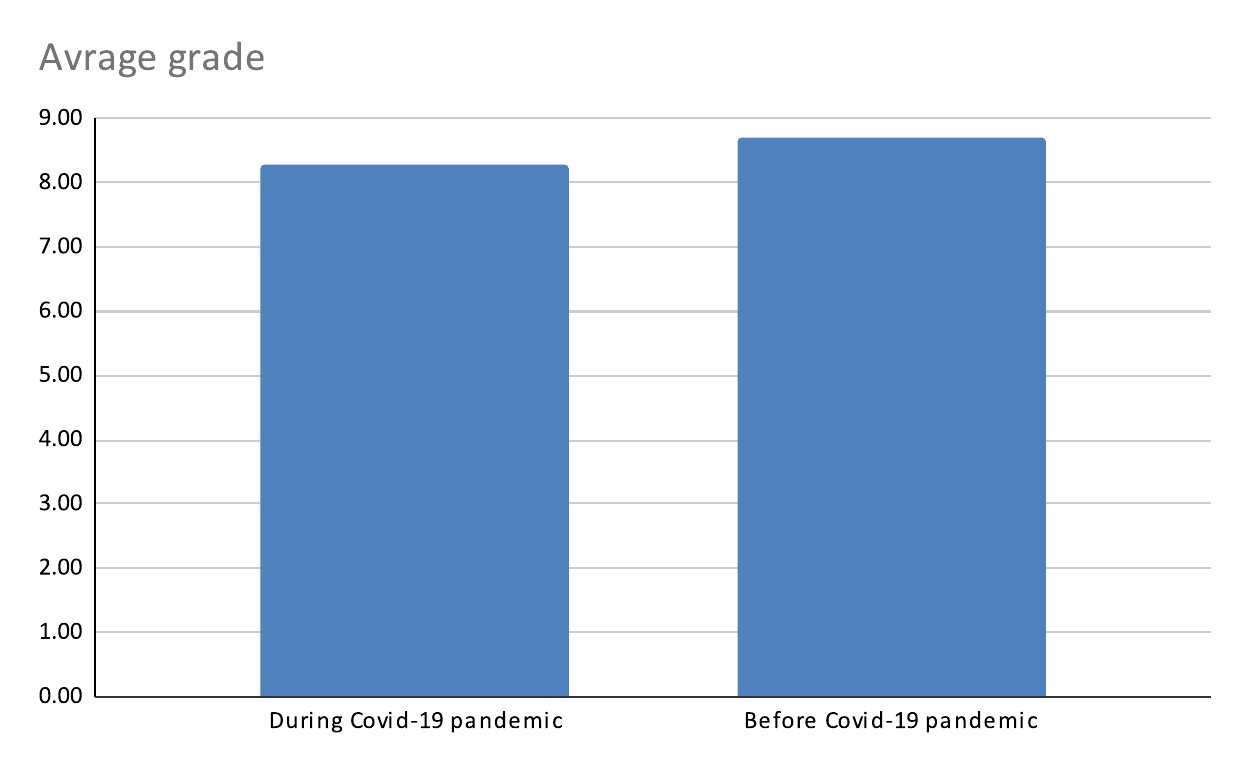}
\caption{Average lab grade of students Before Pandemic (2019) vs. During Pandemic (2020)}
\label{fig:avggrades}
\end{figure}

We can see in Fig. \ref{fig:avggrades} the average lab grade (computed for all the laboratory
tasks, for all students) computed Before Pandemic (2019) and During Pandemic (2020) when we employed Agile techniques. We can see that the average lab grades are very similar, the average grade during pandemic being slightly less than the average grade before pandemic. 
We conclude from this that Agile techniques probably helped in maintaining student grades 
during pandemic (i.e. during online classes) at a level equal to the pre-pandemic one.
These results could also be explained if the students from the 2020 generation are overall better than the students from the 2019 generation. But our analysis of all the grades received at the Web Programming course by the past 5 generations of students did not show
significant differences between these 5 generations of students.

\begin{figure}[h]
\includegraphics [width=0.5\textwidth]{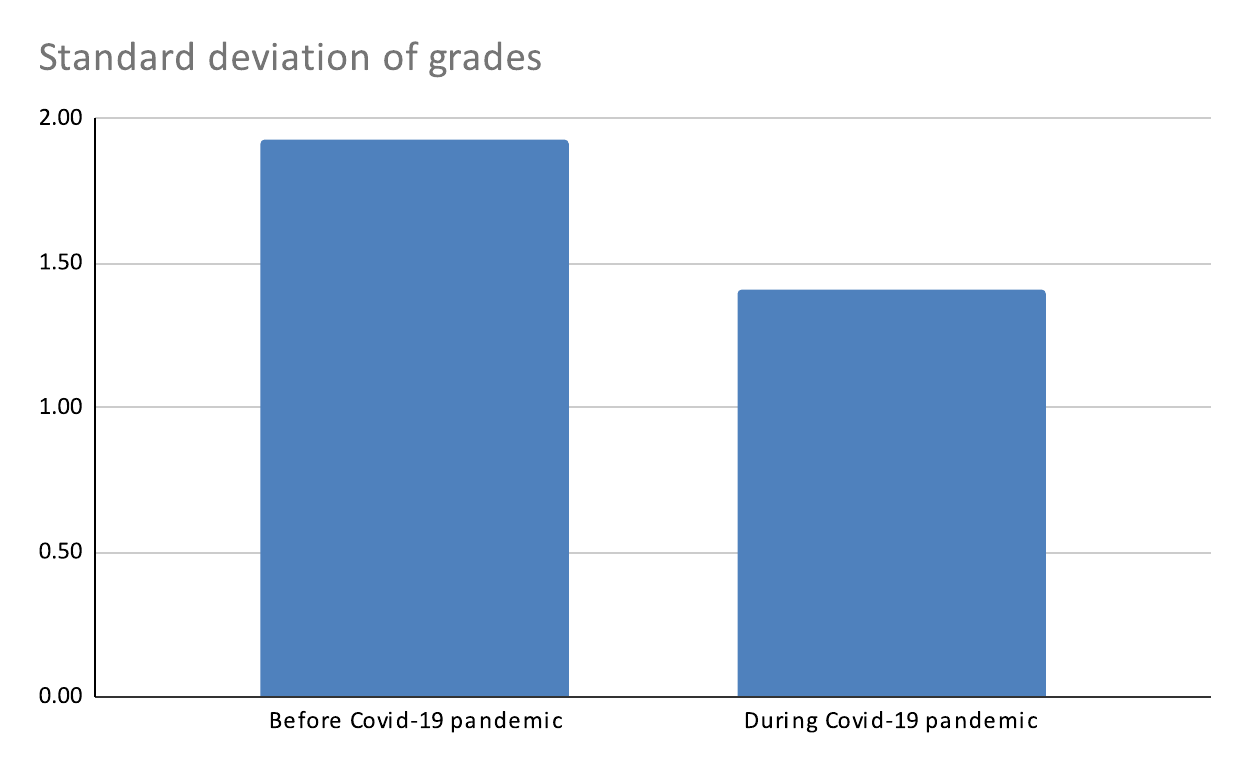}
\caption{The standard deviation of lab grades of students Before Pandemic (2019) vs. During Pandemic (2020)}
\label{fig:stddev}
\end{figure}

\begin{figure}[h]
\includegraphics [width=0.5\textwidth]{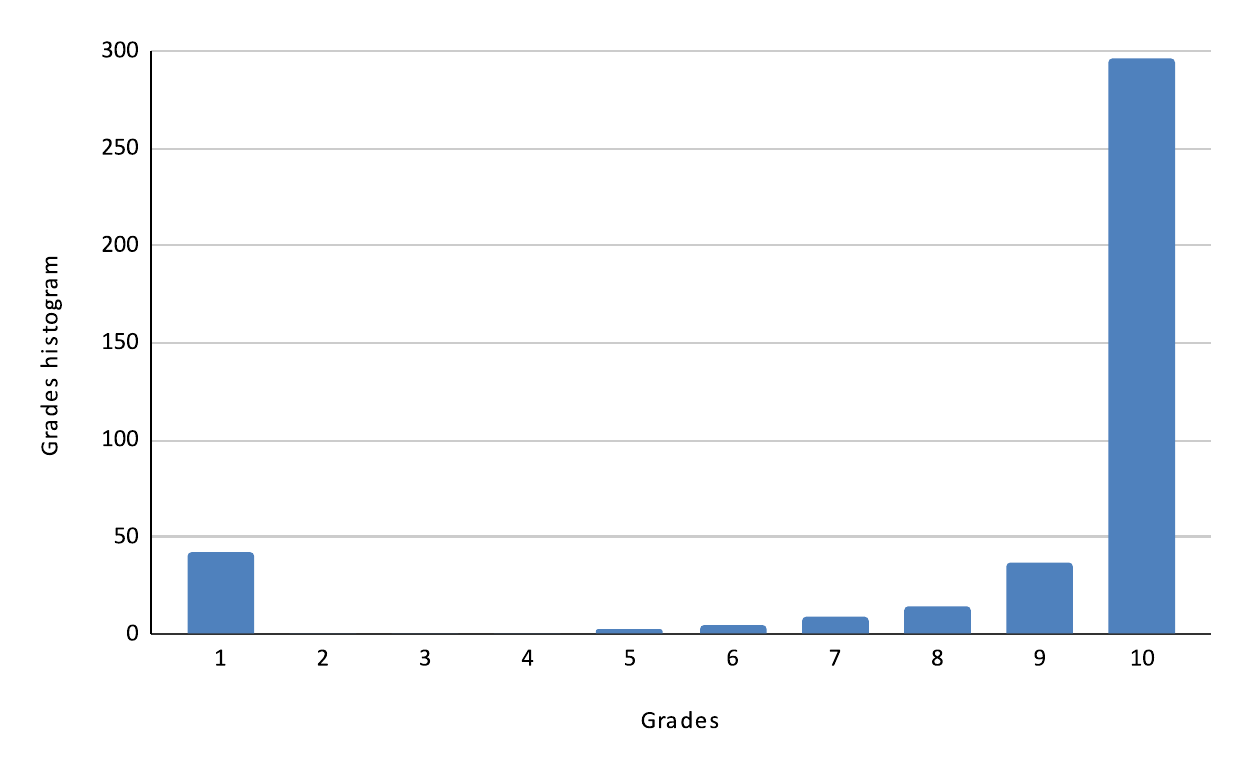}
\caption{The histogram of lab grades (academic year 2019)}
\label{fig:histogram2019}
\end{figure}

\begin{figure}[h]
\includegraphics [width=0.5\textwidth]{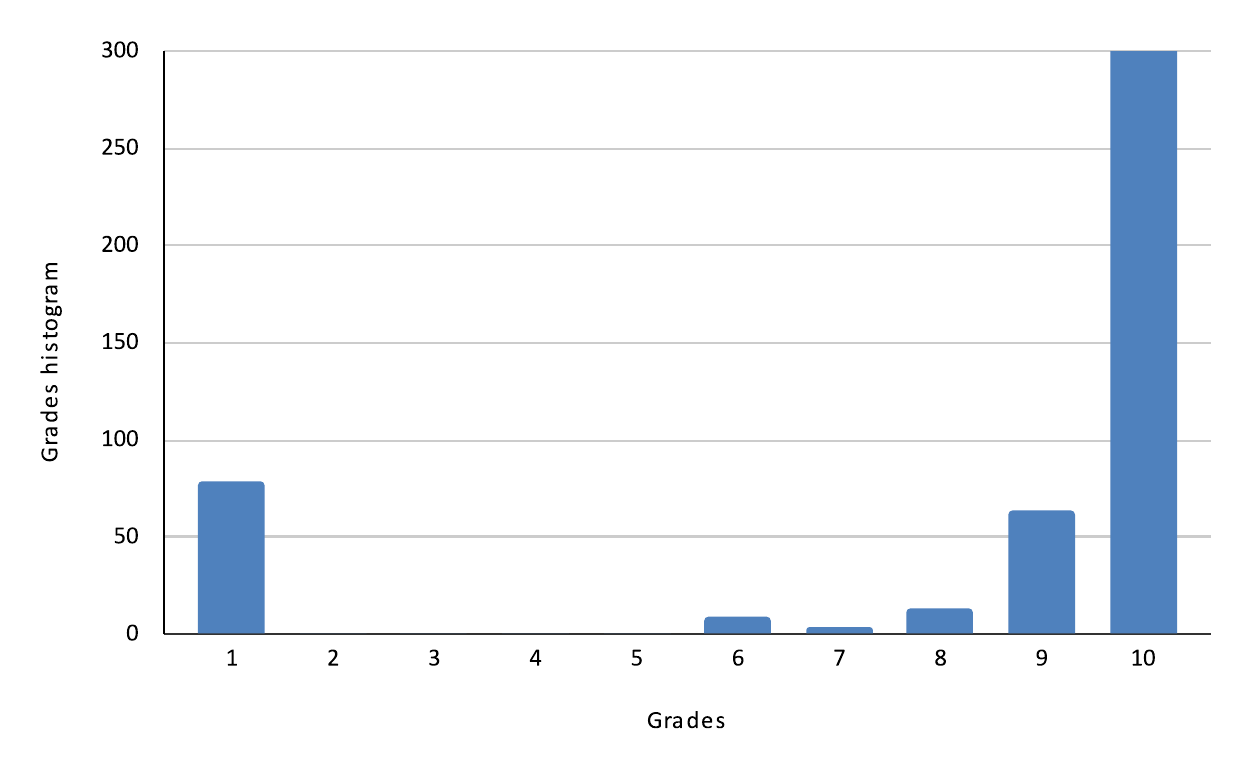}
\caption{The histogram of lab grades (academic year 2020)}
\label{fig:histogram2020}
\end{figure}

Fig. \ref{fig:stddev} shows the standard deviation of the lab grades received by students before pandemic and during pandemic. We can see that the two values are similar. Also,
in Fig. \ref{fig:histogram2019} and \ref{fig:histogram2020} we can see the grades histogram
before the pandemic and during the pandemic, respectively. These plots are very similar, showing that many lab tasks were solved very well by most students.

We can now answer our initial research questions. 
\newline
\emph{Is the Agile methodology helpful in the online courses?}


One of the principal assets of Agile methodology is the principle of systematic retrospective (implemented in the retrospective meetings in software development); the retrospective perspective, repeating ideas and notions is a formidable aid in the learning process, even if is not sufficiently used by the teachers. In the proposed approach, there is time allocated for a retrospective part, thus helping the learners to fix their knowledge. Because if this aspect, using Agile methodology can improve the effectiveness of the online courses. 

\emph{Should Agile methodologies be applied to courses /seminars or laboratory? }

Agile methodology is all about adapting, adapting to obtain better results, to a specific environment or to projects. Parts of the Scrum methodology can not be used for courses if the number of the students exceeds 20-30 people, but the retrospective part can prove itself very useful as it helps achieving and better understand the information presented in the course. 
For the seminars (depending again on the number of students), time can be allocated in the beginning of the seminar to check if there are questions, if the students encountered issues or problems related to the previous taught topic they could not solve or if they found an ingenious method to solve the issues. 
Allocating time in the end of the seminar for a retrospective helps to understand better the concepts and to “place” them in patterns. 
In the laboratories as the student number is the smallest, it’s easier to apply the Agile methodology and to allocate time for the students to speak their mind related to the previous taught topic or the experiment/work they had to do. 
In conclusion, the adapted Agile methodology are useful in the teaching process, starting from the courses, continuing with the seminars and laboratories.

\section{Conclusion and future work}

The experiment of introducing the Agile inspired methodologies in teaching courses, seminars and laboratories (i.e. practical work classes) had benefits and most of the students had a positive reaction to them. However, more than the positive reactions, the grades should reflect if there is an improvement in the learning process and a comparison has to be made with the grades obtained by the students from the previous year. The future work will consist in fine tuning the proposed methodologies (stands-up, retrospective, spring planning) and elaborating some metrics related to these activities (we proposed some time constraints), but we need more studies before we can state that the proposed time intervals are the best ones. 
We have shown by analyzing the grades obtained by students before the pandemic classes and during the pandemic classes that using Agile methodologies in teaching helps the educational process.

\bibliographystyle{apalike}
{\small
\bibliography{example}}

\end{document}